\documentclass[aps,prd,twocolumn,showpacs,nofootinbib]{revtex4-2}

\usepackage{graphicx}
\usepackage{multirow}
\usepackage{bm}
\usepackage{bbm}
\usepackage{braket}
\usepackage{color}
\usepackage{slashed}
\usepackage{amsmath, amssymb}
\usepackage[colorlinks=true,linkcolor=blue,citecolor=blue, urlcolor=blue]{hyperref}

\begin{document}

\title{Reanalysis of exclusive charmonium production at the $B$ factories}

\author{Xu-Dong Huang$^1$}
\email{huangxd@cqnu.edu.cn}
\author{Jian-Xiong Wang$^{2,3}$}
\email{jxwang@ihep.ac.cn}

\affiliation{$^1$ College of Physics and Optoelectronic Engineering, Chongqing Normal University, Chongqing 401331, China}
\affiliation{$^2$ Institute of High Energy Physics, Chinese Academy of Sciences, 19B Yuquan Road, Shijingshan District, Beijing, 100049, P.R. China}
\affiliation{$^3$ University of Chinese Academy of Sciences, Chinese Academy of Sciences, 19A Yuquan Road, Shijingshan District, Beijing, 100049, P.R. China}

\begin{abstract}
In this paper, we present a comprehensive analysis of the next-to-next-to-leading-order (NNLO) QCD corrections to the exclusive processes $e^{+}e^{-}\to J/\psi+\eta_{c}$, $e^{+}e^{-}\to J/\psi+\chi_{cJ}$, $e^{+}e^{-}\to \eta_{c}+\gamma$, and $e^{+}e^{-}\to \chi_{cJ}+\gamma$ $(J=0,1,2)$ at the $B$ factories, comparing perturbative expansions truncated at the amplitude level and squared-amplitude level. Our results show that the amplitude-level prescription improves the perturbative convergence, with the NNLO/LO ratios lying closer to the corresponding NLO/LO ratios than those in the squared-amplitude-level prescription. This improvement is particularly pronounced for $e^{+}e^{-}\to\chi_{c2}+\gamma$, where the NNLO prediction is about 27\% of the LO result in the amplitude-level prescription, compared with only 12\% in the squared-amplitude-level prescription. The amplitude-level prescription also reduces the residual renormalization scale dependence, particularly for the $P$-wave channels. Finally, the NNLO predictions are generally in good agreement with the available Belle and BaBar measurements and have central values closer to the data than those obtained with the squared-amplitude-level prescription. For channels with only upper limits, the predictions from both prescriptions are consistent with the experimental constraints, except for $e^{+}e^{-}\to\eta_c+\gamma$, where both NNLO predictions remain above the current upper limit.
\end{abstract}

\maketitle

\section{Introduction}

In quantum chromodynamics (QCD), the study of heavy quarkonium production plays a crucial role in understanding quark interactions in two-body systems. In processes involving large momentum transfers, perturbative QCD is essential for obtaining theoretical predictions. The exclusive production of charmonium in electron–positron annihilation at $B$ factories has attracted intense theoretical and experimental interest.

Significant experimental measurements of exclusive charmonium production have been performed at the $B$ factories. The $e^+e^- \to J/\psi + \eta_c$ process has been precisely measured by Belle~\cite{Belle:2002tfa, Belle:2004abn} and subsequently confirmed by BaBar~\cite{BaBar:2005nic}. For the $e^+e^- \to J/\psi + J/\psi$ process, only an upper limit has been established by the Belle collaboration~\cite{Belle:2004abn}. In addition, both collaborations have extended their investigations to $P$-wave associated production via $e^+e^- \to J/\psi + \chi_{cJ}~(J=0,1,2)$. Although the $e^+e^- \to J/\psi + \chi_{c0}$ process has been definitively established~\cite{Belle:2004abn, BaBar:2005nic}, the evidence for the $\chi_{c1}$ and $\chi_{c2}$ final states remains inconclusive, yielding only upper limits on their cross sections~\cite{Belle:2004abn}. For the photon-associated processes $e^+e^- \to \eta_c + \gamma$ and $e^+e^- \to \chi_{cJ} + \gamma$ ($J=0,1,2$), the Belle collaboration observed the $e^+e^- \to \chi_{c1} + \gamma$ process and set upper limits on the $\eta_c$, $\chi_{c0}$, and $\chi_{c2}$ channels~\cite{Belle:2018jqa}. Additionally, complementary measurements of the production cross sections for $e^+e^- \to \eta_c/\chi_{cJ} + \gamma$ at lower center-of-mass energies have been performed by the BESIII collaboration~\cite{BESIII:2014uzr, BESIII:2017nty, BESIII:2021yal}.

These experimental findings have stimulated extensive theoretical efforts, mostly carried out within the framework of nonrelativistic QCD (NRQCD) factorization~\cite{Bodwin:1994jh}. For the processes $e^+e^- \to J/\psi + \eta_c$ and $e^+e^- \to J/\psi + \chi_{cJ}$, the production cross sections have been calculated up to the two-loop level~\cite{Braaten:2002fi, Liu:2002wq, Hagiwara:2003cw, Bodwin:2006ke, He:2007te, Bodwin:2007ga, Zhang:2005cha, Gong:2007db, Dong:2012xx, Li:2013otv, Feng:2019zmt, Huang:2022dfw, Li:2025mng, Chen:2025qgy, Maxia:2026fbz, Zhang:2008gp, Wang:2011qg, Dong:2011fb, Wang:2013vn, Jiang:2018wmv, Sun:2021tma, Sang:2022kub, Liu:2026ray}, and the results exhibit remarkable agreement with the experimental measurements. For the $e^{+}e^{-}\to J/\psi+J/\psi$ process, the production cross section has been calculated up to the two-loop level~\cite{Braaten:2002fi, Bodwin:2002kk,Bodwin:2002fk, Davier:2006fu, Bodwin:2006yd, Gong:2008ce, Braguta:2008tg, Fan:2012dy, Sang:2023liy, Huang:2023pmn, Chen:2026maw, Maxia:2026fbz}, and the predicted result is relatively small and consistent with the Belle upper limit. For the $e^{+}e^{-}\to\eta_{c}+\gamma$ and $e^{+}e^{-}\to\chi_{cJ}+\gamma$ processes, the production cross sections have been calculated up to the two-loop level~\cite{Chung:2008km, Sang:2009jc, Li:2009ki, Fan:2012dy, Wang:2013ywc, Xu:2014zra, Brambilla:2017kgw, Chen:2017pyi, Chung:2019ota, Yu:2020tri, Sang:2020fql, Li:2025pbt}, and provide complementary and valuable tests of the NRQCD factorization formalism.

The combination of high-precision theoretical predictions and experimental measurements provides a solid foundation for detailed phenomenological studies of these processes. While next-to-next-to-leading order (NNLO) calculations generally yield reliable theoretical predictions for these processes, the perturbative truncation deserves particular attention when higher-order corrections become large. It has been found that the NNLO QCD correction for the $e^+e^- \to J/\psi + J/\psi$ process can be substantially negative, leading to poor perturbative convergence and even an unphysical negative prediction for the production cross section~\cite{Huang:2023pmn, Sang:2023liy}. These observations highlight the importance of a consistent prescription for truncating the perturbative series. Our previous studies for the processes $e^+e^- \to J/\psi + J/\psi$ and $J/\psi \to 3\gamma$ demonstrate that amplitude-level truncation leads to reduced renormalization-scale dependence and improved perturbative convergence compared with squared-amplitude-level truncation~\cite{Huang:2023pmn, Zeng:2026ois}. Then, it is natural to investigate the phenomenological impact of fixed-order truncation at the amplitude and squared-amplitude levels in other exclusive production processes. In this work, we provide a comprehensive analysis of the NNLO QCD corrections to the exclusive processes $e^+e^- \to J/\psi + \eta_c$, $e^+e^- \to J/\psi + \chi_{cJ}$, $e^+e^- \to \eta_c + \gamma$, and $e^{+}e^{-}\to \chi_{cJ}+ \gamma$ at $B$ factories.

\section{NRQCD Factorization}

Within the NRQCD factorization framework, the cross sections for the exclusive processes $e^{+}e^{-}\to J/\psi+\eta_{c}$, $e^{+}e^{-}\to J/\psi+\chi_{cJ}$, $e^{+}e^{-}\to \eta_{c}+\gamma$, and $e^{+}e^{-}\to \chi_{cJ}+\gamma$ can be written as:
\begin{eqnarray}
&&\frac{d\sigma[e^+ e^- \to X_1(\lambda_1) + X_2(\lambda_2)]}{d \cos\theta} \nonumber\\
&&=\frac{2\pi\alpha}{s^{\frac{3}{2}}}\frac{d\Gamma[\gamma^*\to X_1(\lambda_1)+X_2(\lambda_2)]}{d \cos\theta} 
\nonumber\\
&&=\frac{\alpha}{8s^2}\left(\frac{|\mathbf{P}_1|}{\sqrt{s} }\right)|\mathcal{A}^{X_1, X_2}_{\lambda_1 , \lambda_2}|^2 \times \big\{^{\frac{1+\cos^2 \theta}{2}, ~~~     \lambda=\pm 1}_{1-\cos^2 \theta,  ~~~~   \lambda=0}, \label{dsigma}
\end{eqnarray}
where the pair $(X_1,X_2)$ takes the values $(J/\psi,\eta_c)$, $(J/\psi,\chi_{cJ})$, $(\eta_c,\gamma)$, or $(\chi_{cJ},\gamma)$; $\alpha$ denotes the QED fine-structure constant, $\sqrt{s}$ is the center-of-mass (CM) energy, $\theta$ is the polar angle between the outgoing $X_1$ and the electron beam direction, and $\lambda_1$ and $\lambda_2$ are the helicities of $X_1$ and $X_2$, respectively. The quantity $\mathcal{A}^{X_1,X_2}_{\lambda_1,\lambda_2}$ denotes the corresponding helicity amplitude, and $\lambda\equiv\lambda_1-\lambda_2$. Angular-momentum conservation requires $|\lambda|\leq 1$. The magnitude of the three-momentum of $X_1$ is given by
\begin{eqnarray}
|\mathbf{P}_1|=\frac{\lambda^{\frac{1}{2}}(s,M_{X_1}^2, M_{X_2}^2)}{2\sqrt{s}},
\end{eqnarray}
where $\lambda(x,y, z)=x^2+y^2+z^2-2(xy+xz+yz).$

Integrating Eq.~(\ref{dsigma}) over $\theta$ and summing over all possible helicity states of $X_1$ and $X_2$, we obtain the total unpolarized cross section:
\begin{eqnarray}
\sigma(J/\psi+\eta_c)&=&\frac{\alpha}{6s^2}\frac{\lvert\mathbf{P}_1\rvert}{\sqrt{s}}2\lvert\mathcal{A}^{J/\psi,\eta_c}_{1,0}\rvert^2,\nonumber\\
\sigma(J/\psi+\chi_{c0})&=&\frac{\alpha}{6s^2}\frac{\lvert\mathbf{P}_1\rvert}{\sqrt{s}}\bigg(2\lvert\mathcal{A}^{J/\psi,\chi_{c0}}_{1,0}\rvert^2+\lvert\mathcal{A}^{J/\psi,\chi_{c0}}_{0,0}\rvert^2\bigg),\nonumber\\
\sigma(J/\psi+\chi_{c1})&=&\frac{\alpha}{6s^2}\frac{\lvert\mathbf{P}_1\rvert}{\sqrt{s}}\bigg(2\lvert\mathcal{A}^{J/\psi,\chi_{c1}}_{1,0}\rvert^2+2\lvert\mathcal{A}^{J/\psi,\chi_{c1}}_{0,1}\rvert^2\nonumber\\ &&+2\lvert\mathcal{A}^{J/\psi,\chi_{c1}}_{1,1}\rvert^2\bigg),\nonumber\\
\sigma(J/\psi+\chi_{c2})&=&\frac{\alpha}{6s^2}\frac{\lvert\mathbf{P}_1\rvert}{\sqrt{s}}\bigg(2\lvert\mathcal{A}^{J/\psi,\chi_{c2}}_{1,0}\rvert^2+2\lvert\mathcal{A}^{J/\psi,\chi_{c2}}_{0,1}\rvert^2\nonumber\\
&&\hspace{-1cm}+2\lvert\mathcal{A}^{J/\psi,\chi_{c2}}_{1,1}\rvert^2+2\lvert\mathcal{A}^{J/\psi,\chi_{c2}}_{1,2}\rvert^2+\lvert\mathcal{A}^{J/\psi,\chi_{c2}}_{0,0}\rvert^2\bigg),\nonumber\\
\sigma(\eta_c+\gamma)&=&\frac{\alpha}{6s^2}\left(\frac{|\mathbf{P}_1|}{\sqrt{s} }\right)2|\mathcal{A}^{\eta_c,\gamma}_{0 , 1}|^2,\nonumber\\
\sigma(\chi_{c0}+\gamma)&=&\frac{\alpha}{6s^2}\left(\frac{|\mathbf{P}_1|}{\sqrt{s} }\right)2|\mathcal{A}^{\chi_{c0},\gamma}_{0 , 1}|^2,\nonumber\\
\sigma(\chi_{c1}+\gamma)&=&\frac{\alpha}{6s^2}\left(\frac{|\mathbf{P}_1|}{\sqrt{s} }\right)(2|\mathcal{A}^{\chi_{c1},\gamma}_{1 , 1}|^2+2|\mathcal{A}^{\chi_{c1},\gamma}_{0 , 1}|^2),\nonumber\\
\sigma(\chi_{c2}+\gamma)&=&\frac{\alpha}{6s^2}\left(\frac{|\mathbf{P}_1|}{\sqrt{s} }\right)(2|\mathcal{A}^{\chi_{c2},\gamma}_{2 ,1}|^2+2|\mathcal{A}^{\chi_{c2},\gamma}_{1 , 1}|^2\nonumber\\
&&+2|\mathcal{A}^{\chi_{c2},\gamma}_{0 , 1}|^2).  \label{sigma0}
\end{eqnarray}

For hard exclusive reactions involving quarkonium, NRQCD factorization can be formulated directly at the amplitude level~\footnote{For exclusive processes, all IR divergences already cancel at the amplitude level.}. Specifically, the helicity amplitudes $\mathcal{A}^{X_1,X_2}_{\lambda_1,\lambda_2}$ can be written as
\begin{eqnarray}
&&\mathcal{A}^{J/\psi,\eta_c}_{\lambda_1 , \lambda_2}=c^{J/\psi,\eta_c}_{\lambda_1 , \lambda_2} \sqrt{2M_{J/\psi}}\sqrt{2M_{\eta_c}}\frac{\langle \mathcal{O}_{J/\psi}\rangle \langle \mathcal{O}_{\eta_c}\rangle}{m_c^{2}}, \nonumber \\
&&\mathcal{A}^{J/\psi,\chi_{cJ}}_{\lambda_1 , \lambda_2}=c^{J/\psi,\chi_{cJ}}_{\lambda_1 , \lambda_2} \sqrt{2M_{J/\psi}}\sqrt{2M_{\chi_{cJ}}}\frac{\langle \mathcal{O}_{J/\psi}\rangle \langle \mathcal{O}_{\chi_{cJ}}\rangle}{m_c^{4}}, \nonumber \\
&&\mathcal{A}^{\eta_c,\gamma}_{\lambda_1 , \lambda_2}=c^{\eta_c,\gamma}_{\lambda_1 , \lambda_2} \sqrt{2M_{\eta_c}}\frac{\langle \mathcal{O}_{\eta_c}\rangle}{m_c}, \nonumber \\
&&\mathcal{A}^{\chi_{cJ},\gamma}_{\lambda_1 , \lambda_2}=c^{\chi_{cJ},\gamma}_{\lambda_1 , \lambda_2} \sqrt{2M_{\chi_{cJ}}}\frac{\langle \mathcal{O}_{\chi_{cJ}}\rangle}{m_c^{2}}, \label{nrqcd-formula}
\end{eqnarray}
where the short-distance coefficients (SDCs) $c^{X_1,X_2}_{\lambda_1,\lambda_2}$ contain the perturbatively calculable short-distance contributions. The SDCs for $e^{+}e^{-}\to J/\psi+\eta_c$ can be obtained from the electromagnetic form factor $F(s)$ given in Refs.~\cite{Feng:2019zmt, Li:2025mng}, since the helicity amplitude satisfies $\mathcal{A}^{J/\psi,\eta_c}_{1,0}=e\sqrt{s}|\mathbf{P}_1|F(s)$. The explicit expressions for $F(s)$ and $c^{X_1, X_2}_{\lambda_1 , \lambda_2}$ up to two-loop order can be found in Refs.~\cite{Li:2025mng, Li:2025pbt, Liu:2026ray}. The prefactors $\sqrt{2M_{J/\psi}}$, $\sqrt{2M_{\eta_c}}$, and $\sqrt{2M_{\chi_{cJ}}}$ arise because relativistic normalization is employed for the quarkonium states in the helicity amplitudes, whereas the long-distance matrix elements (LDMEs) are defined using nonrelativistic normalization. The NRQCD LDMEs are defined as
\begin{eqnarray}
&&\langle\mathcal{O}_{\eta_c}\rangle=\langle \eta_c|\psi^\dagger\chi(\mu_\Lambda)|0\rangle,\nonumber\\
&&\langle\mathcal{O}_{J/\psi}\rangle=\langle J/\psi|\psi^\dagger {\mathbf \sigma} \cdot \epsilon_{J/\psi} \chi(\mu_\Lambda)|0\rangle,\nonumber\\
&&\langle\mathcal{O}_{\chi_{c0}}\rangle=\langle \chi_{c0}|\psi^\dagger\frac{1}{\sqrt{3}}\left (-\frac{i}{2}\overleftrightarrow{D}\cdot\sigma \right )\chi(\mu_\Lambda)|0\rangle,\nonumber\\
&&\langle\mathcal{O}_{\chi_{c1}}\rangle=\langle \chi_{c1}|\psi^\dagger\frac{1}{\sqrt{2}}\left (-\frac{i}{2}\overleftrightarrow{D}\times\sigma\right )\cdot\epsilon_{\chi_{c1}}\chi(\mu_\Lambda)|0\rangle,\nonumber\\
&&\langle\mathcal{O}_{\chi_{c2}}\rangle=\langle \chi_{c2}|\psi^\dagger\left( -\frac{i}{2}\overleftrightarrow{D}^{(i}\sigma^{j)}\epsilon^{ij}_{\chi_{c2}}\right)\chi(\mu_\Lambda)|0\rangle, \label{ld}
\end{eqnarray}
where $\epsilon_{J/\psi}$ and $\epsilon_{\chi_{c1}}$ denote the polarization vectors of $J/\psi$ and $\chi_{c1}$, respectively, and $\epsilon_{\chi_{c2}}$ denotes the polarization tensor of $\chi_{c2}$. Here, the factorization scale $\mu_\Lambda$ is expected to cancel against the corresponding scale dependence of the SDCs.

\section{Truncation Prescription}

In the following, we introduce two different prescriptions for truncating the perturbative expansion of the exclusive charmonium production cross sections. The amplitude is expanded in powers of the strong coupling $\alpha_s$ as
\begin{eqnarray}
\mathcal{A} = \mathcal{A}^{(0)} + \mathcal{A}^{(1)} +\mathcal{A}^{(2)} + \cdots,
\end{eqnarray} 
where $\mathcal{A}^{(i)}$ with $i=0,1,2$ denote the tree-level, one-loop, and two-loop amplitudes, respectively. For convenience, we denote the power of $\alpha_s$ in $\mathcal{A}^{(0)}$ by $n_0$.

In the squared-amplitude-level prescription, one first constructs the full squared amplitude, $|\mathcal{A}|^2 = \sum_{i,j} \mathcal{A}^{(i)*}\mathcal{A}^{(j)}$, and organizes its terms according to their total power of $\alpha_s$. The resulting series is then truncated at a fixed order $n$ (e.g., $n=2n_0+2$ for NNLO):
\begin{eqnarray}
\sigma_{\mathrm{sq}} \propto \sum_{k=2n_0}^{n} \left(|\mathcal{A}|^2\right)^{(k)}, 
\left(|\mathcal{A}|^2\right)^{(k)} = \sum_{i+j=k} \mathcal{A}^{(i)*}\mathcal{A}^{(j)}.
\end{eqnarray} 
This prescription retains only terms of order $\alpha_s^k$ with $k\leq n$ and discards all higher-order contributions, even if they arise from squaring the truncated amplitude.

In the amplitude-level prescription, one first truncates the amplitude at a fixed order $n$ and then squares the truncated amplitude~\footnote{This prescription has been applied to the leptonic decay of quarkonium~\cite{Beneke:1997jm, Feng:2022vvk}.}. That is,
\begin{eqnarray}
\sigma_{\mathrm{amp}} \propto \left|\mathcal{A}_{\mathrm{amp}}\right|^2, 
\mathcal{A}_{\mathrm{amp}} = \sum_{i=0}^{n} \mathcal{A}^{(i)}.
\end{eqnarray} 
This prescription includes all contributions up to $\mathcal{O}(\alpha_s^n)$ in the squared amplitude, while also retaining higher-order cross terms, such as $\mathcal{A}^{(i)*}\mathcal{A}^{(j)}$ with $i+j>n$, which are systematically discarded in the squared-amplitude-level prescription.

The difference between the two prescriptions is therefore most transparent in their treatment of higher-order terms. The amplitude-level prescription automatically retains certain partial higher-order contributions. For example, when the amplitude is truncated at NNLO, the interference of the NLO and NNLO amplitudes and the square of the NNLO amplitude generate contributions of $\mathcal{O}(\alpha_s^{n+1})$ and $\mathcal{O}(\alpha_s^{n+2})$, respectively, which are retained in the amplitude-level prescription but omitted in the squared-amplitude-level prescription. The numerical impact of these additional higher-order contributions will be examined in the following analysis.

\section{Phenomenological results} \label{II}

Before presenting the phenomenological predictions, we specify the input parameters used in our analysis. The charm-quark mass is taken to be $m_c = 1.5 \pm 0.2$ GeV. The center-of-mass energy is set to $\sqrt{s} = 10.58~\mathrm{GeV}$, and the QED coupling is taken as $\alpha(\sqrt{s}) = 1/130.9$. We use the package \texttt{RunDec3}~\cite{Herren:2017osy} to evaluate the running QCD coupling $\alpha_s(\mu_R)$. The quarkonium masses are set to their physical values as given by the Particle Data Group~\cite{ParticleDataGroup:2026mpi}:
\begin{eqnarray}
&&M_{\eta_c}=2.9798~\mathrm{GeV},\nonumber\\
&&M_{J/\psi}=3.0969~\mathrm{GeV},\nonumber\\
&&M_{\chi_{c0}}=3.41471~\mathrm{GeV},\nonumber\\
&&M_{\chi_{c1}}=3.51067~\mathrm{GeV},\nonumber\\
&&M_{\chi_{c2}}=3.55617~\mathrm{GeV}.
\end{eqnarray}

The NRQCD factorization scale is fixed at $\mu_\Lambda = 1$ GeV. The corresponding LDMEs at this scale are taken from Refs.~\cite{Chung:2008km,Bodwin:2007fz,Bodwin:2006dn}:
\begin{eqnarray}
&&|\langle{\mathcal{O}_{\eta_c}}\rangle(1\,{\rm GeV})|^2=0.437~\mathrm{GeV}^3,\nonumber\\
&&|\langle{\mathcal{O}_{J/\psi}}\rangle(1\,{\rm GeV})|^2=0.440~\mathrm{GeV}^3,\nonumber\\
&&|\langle{\mathcal{O}_{\chi_{c0}}} \rangle(1\,{\rm GeV})|^2=0.051~\mathrm{GeV}^5,\nonumber\\
&&|\langle{\mathcal{O}_{\chi_{c1}}} \rangle(1\,{\rm GeV})|^2=0.060~\mathrm{GeV}^5,\nonumber\\
&&|\langle{\mathcal{O}_{\chi_{c2}}} \rangle(1\,{\rm GeV})|^2=0.068~\mathrm{GeV}^5.
\end{eqnarray}

\begin{figure*}[htbp]
\centering
\includegraphics[width=0.45\textwidth]{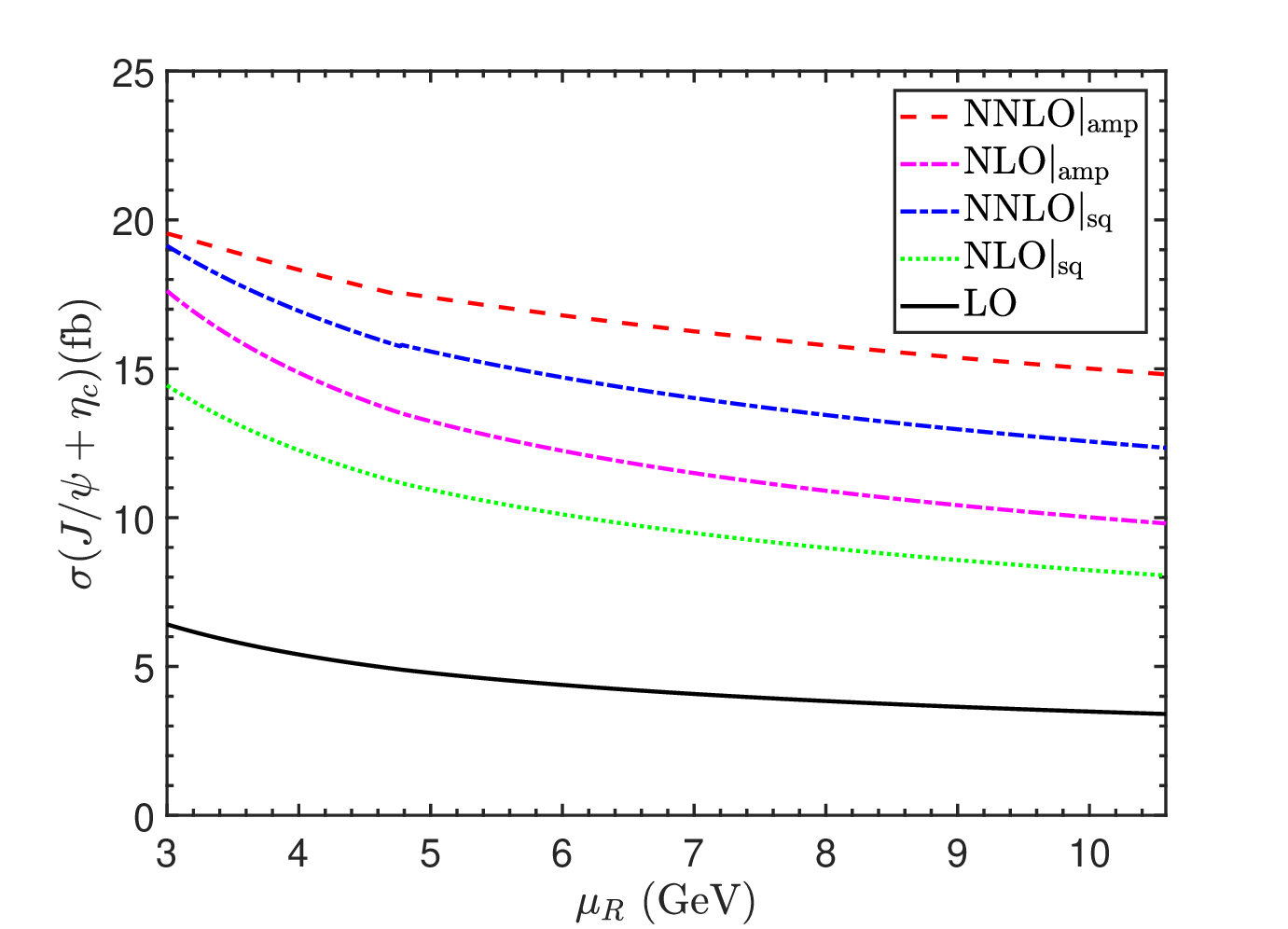}
\includegraphics[width=0.45\textwidth]{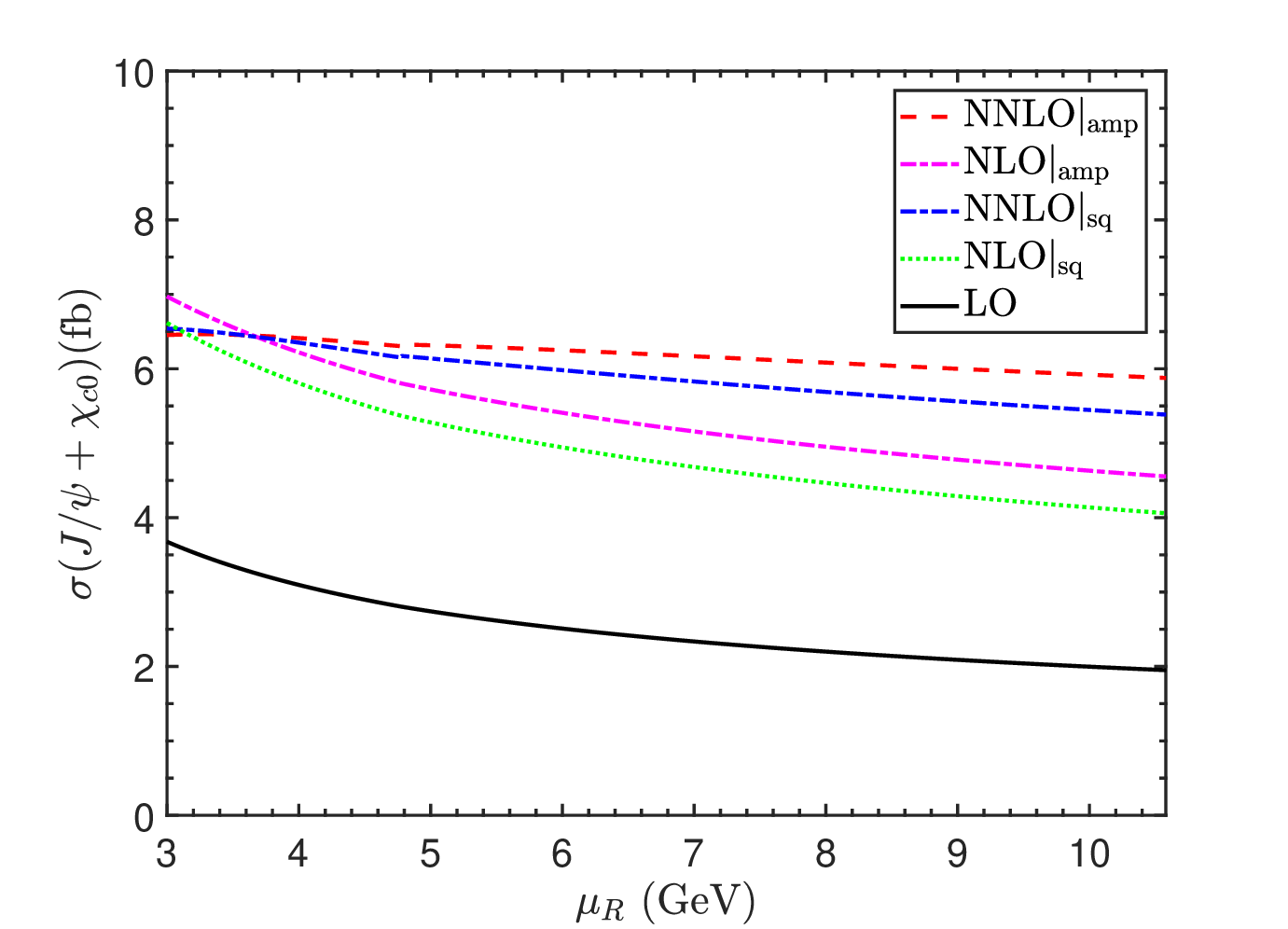}
\includegraphics[width=0.45\textwidth]{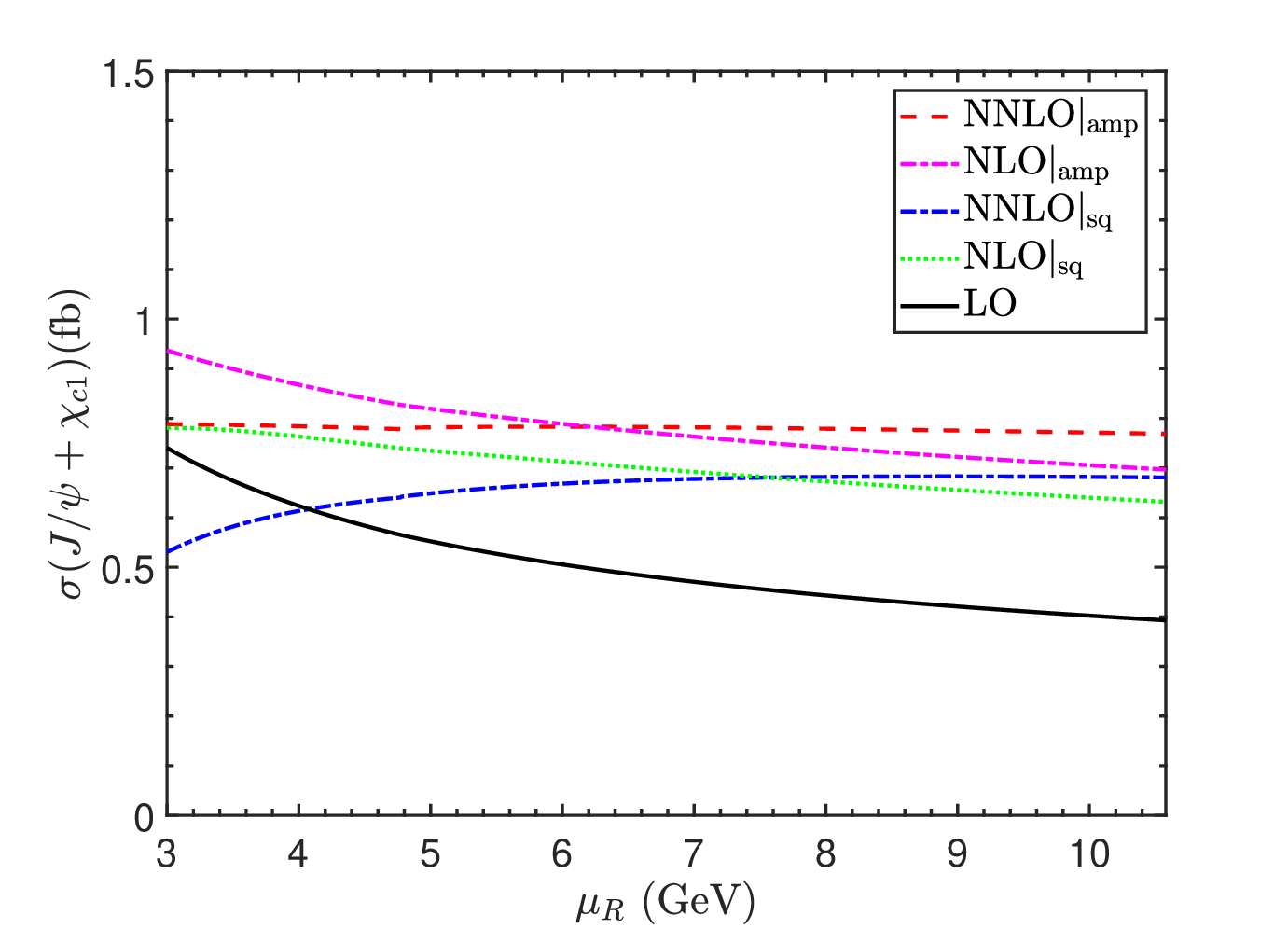}
\includegraphics[width=0.45\textwidth]{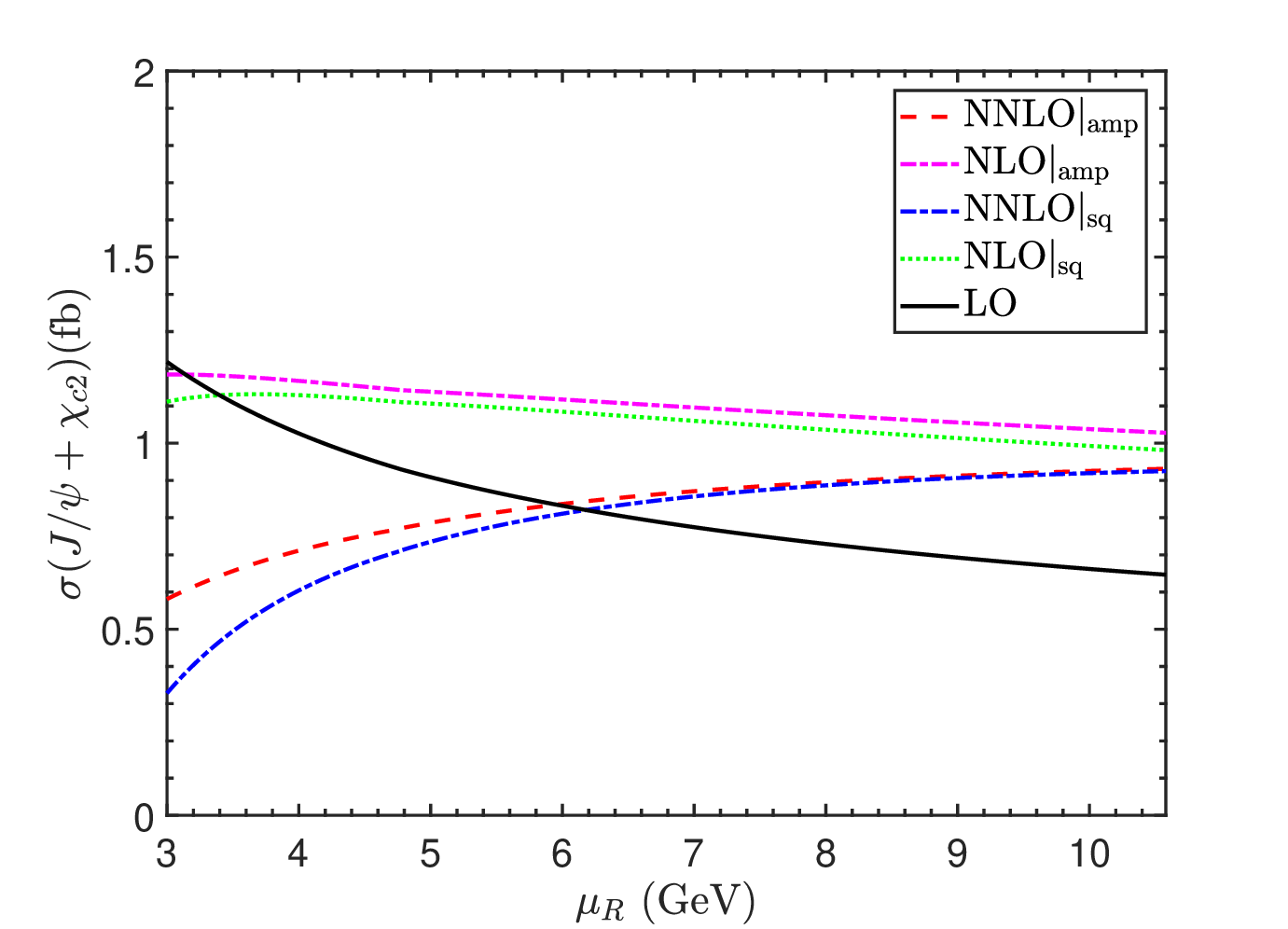}
\includegraphics[width=0.45\textwidth]{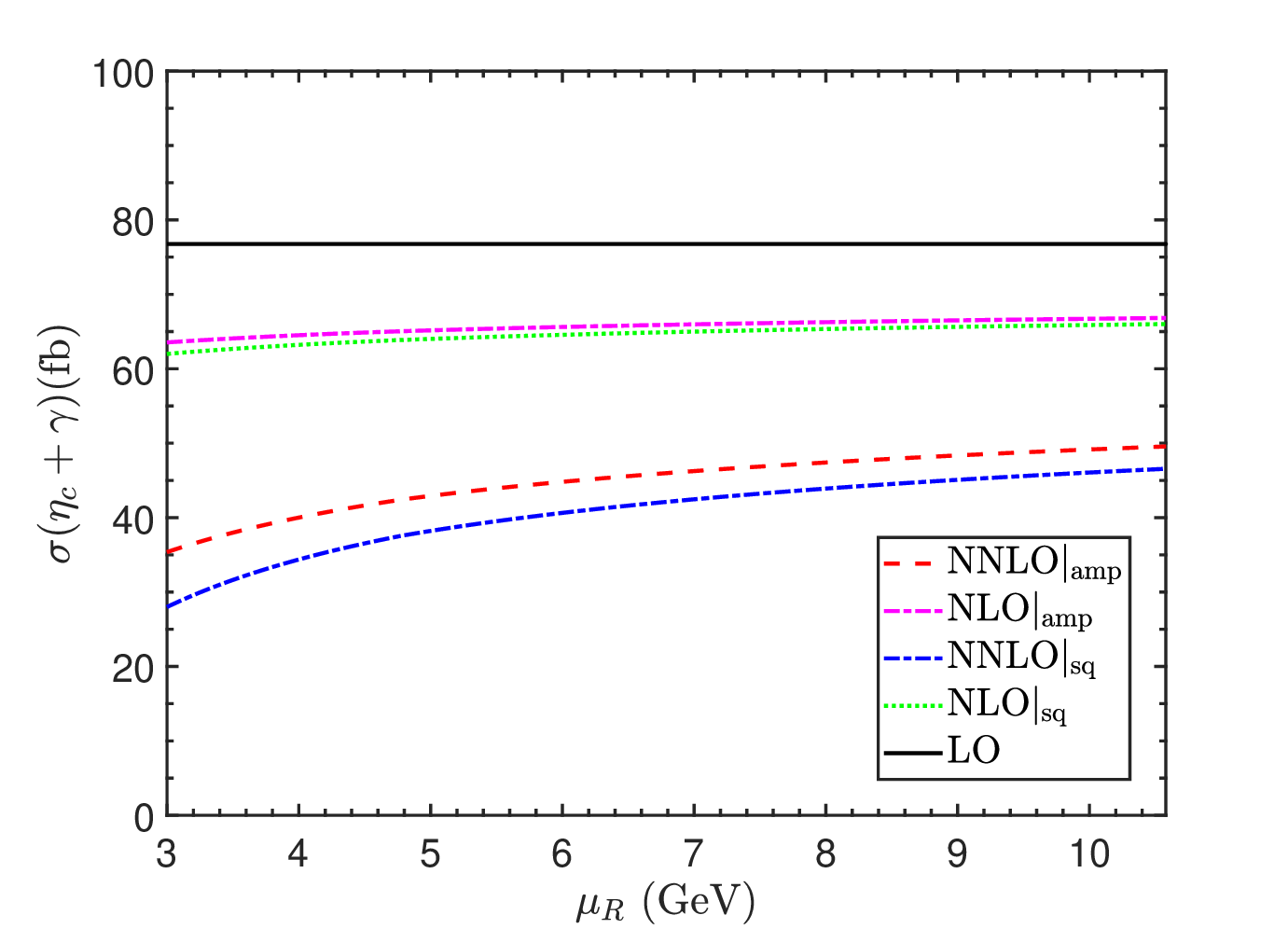}
\includegraphics[width=0.45\textwidth]{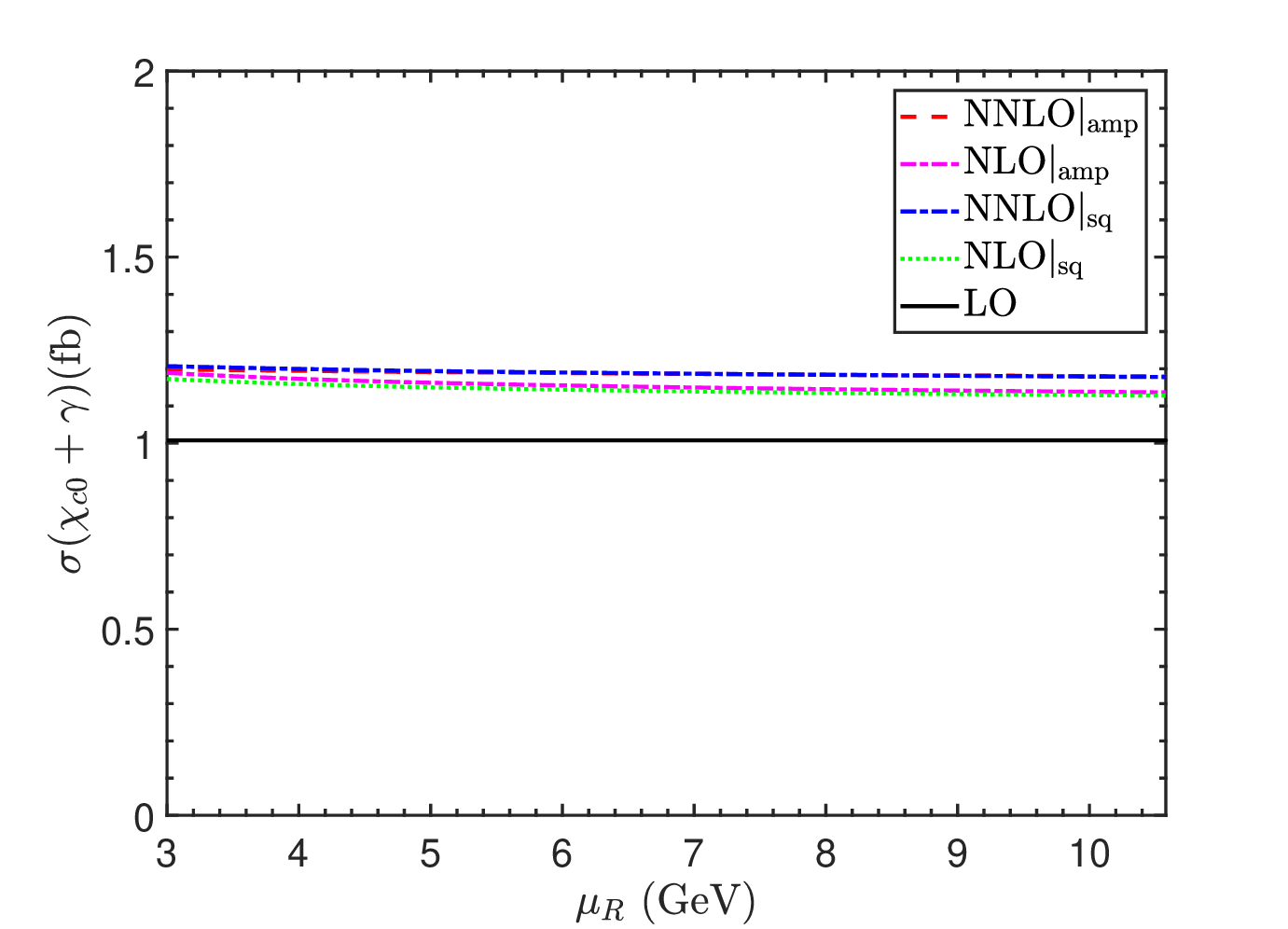}
\includegraphics[width=0.45\textwidth]{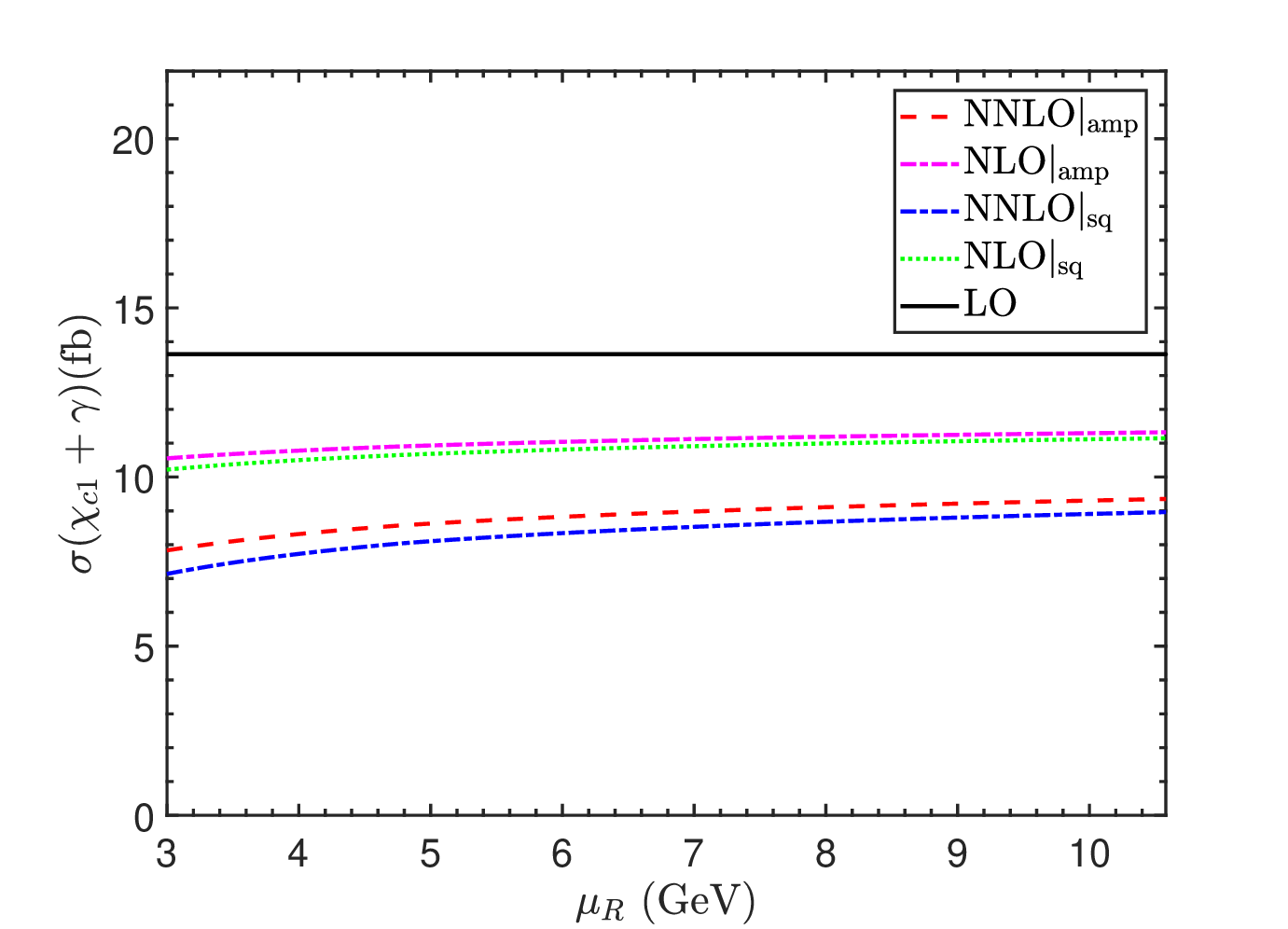}
\includegraphics[width=0.45\textwidth]{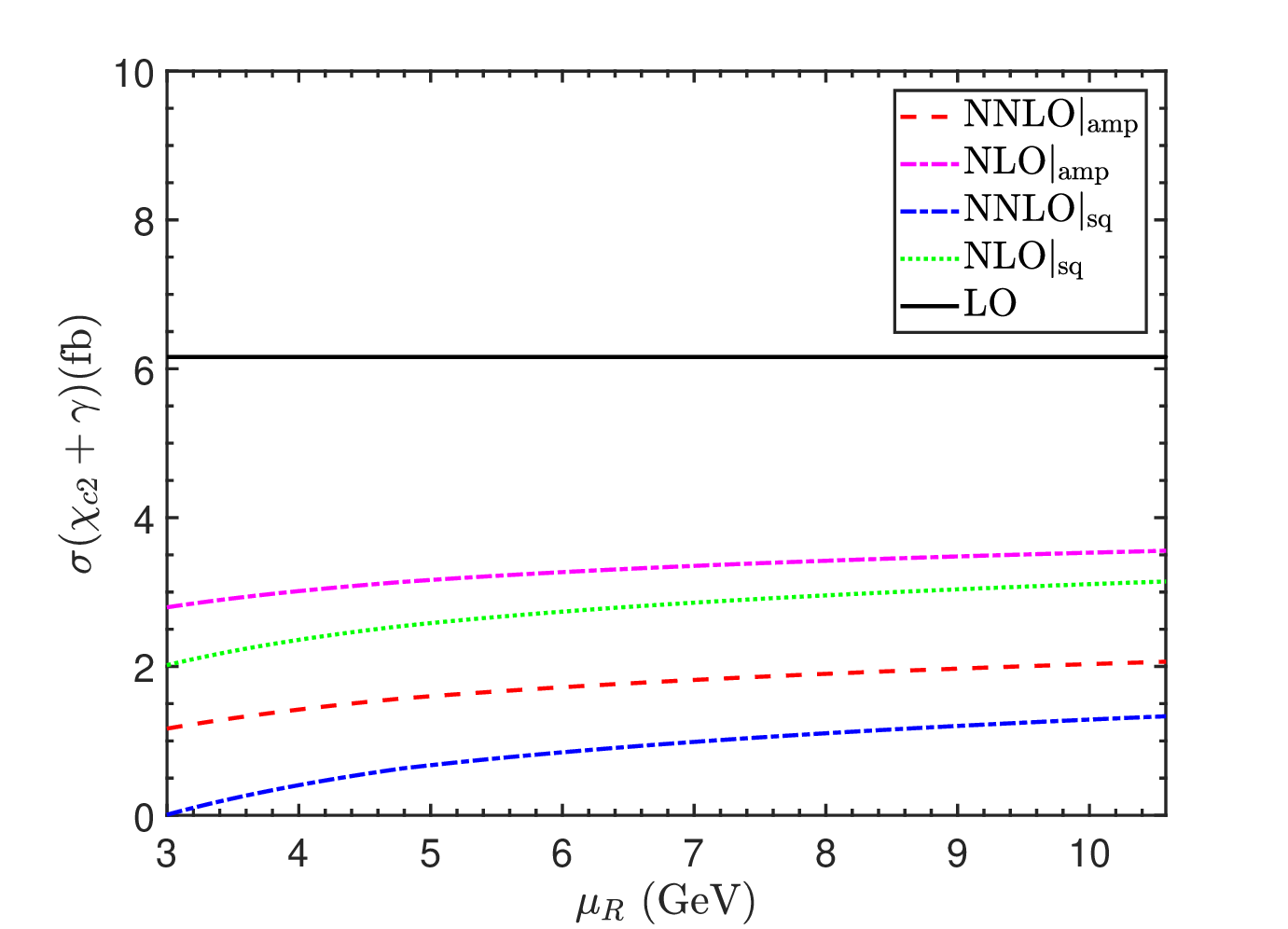}
\caption{The $\mu_R$ dependence of the cross sections for $e^{+}e^{-}\to J/\psi+\eta_c$, $e^{+}e^{-}\to J/\psi+\chi_{cJ}$, $e^{+}e^{-}\to\eta_c+\gamma$, and $e^{+}e^{-}\to\chi_{cJ}+\gamma$ $(J=0,1,2)$ at various perturbative orders.} \label{urdependence}
\end{figure*}

In Fig.~\ref{urdependence}, we show the dependence of the cross sections on the renormalization scale $\mu_R$ for these processes at various perturbative orders. The figure contains eight panels, each corresponding to one of the channels under consideration. The leading-order (LO), next-to-leading-order (NLO), and NNLO results obtained with the squared-amplitude-level and amplitude-level prescriptions are displayed together for comparison.

It can be seen that the NNLO predictions in the squared-amplitude-level prescription, denoted by NNLO$|_{\rm sq}$, exhibit a larger $\mu_R$ dependence than those in the amplitude-level prescription, denoted by NNLO$|_{\rm amp}$. This enhanced scale dependence is particularly pronounced for the channels involving $P$-wave charmonia, such as $e^{+}e^{-}\to J/\psi+\chi_{cJ}$ and $e^{+}e^{-}\to\chi_{cJ}+\gamma$, and is most significant for $e^{+}e^{-}\to\chi_{c2}+\gamma$. In contrast, the $\mu_R$ dependence of the NNLO predictions is reduced in the amplitude-level prescription across all channels. 

\begin{table}[!h]
\begin{center}
\caption{Cross sections (in units of fb) for $e^{+}e^{-}\to J/\psi+\eta_{c}$, $e^{+}e^{-}\to J/\psi+\chi_{cJ}$, $e^{+}e^{-}\to \eta_{c}+\gamma$, and $e^{+}e^{-}\to \chi_{cJ}+\gamma$ $(J=0,1,2)$ up to NNLO in the squared-amplitude-level and amplitude-level prescriptions. The uncertainties arise from varying the renormalization scale $\mu_R$ between $3$ GeV and $\sqrt{s}$, with the central value set to $\sqrt{s}/2$.}
\label{tab:cross:section1}
\resizebox{\columnwidth}{!}{
\begin{tabular}{|c|c|c|c|c|c|}
\hline
 & LO & NLO$|_{\rm sq}$ & NNLO$|_{\rm sq}$ & NLO$|_{\rm amp}$ & NNLO$|_{\rm amp}$\\ \hline
$\sigma(J/\psi+\eta_{c})$ & $4.65^{+1.76}_{-1.25}$ & $10.67^{+3.77}_{-2.61}$ & $15.31^{+3.82}_{-2.96}$ & $12.92^{+4.70}_{-3.11}$ & $17.22^{+2.33}_{-2.40}$ \\  
$\sigma(J/\psi+\chi_{c0})$ & $2.67^{+1.01}_{-0.72}$ & $5.17^{+1.44}_{-1.11}$ & $6.09^{+0.45}_{-0.71}$ & $5.62^{+1.35}_{-1.07}$ & $6.30^{+0.15}_{-0.42}$ \\  
$\sigma(J/\psi+\chi_{c1})$ & $0.54^{+0.20}_{-0.14}$ & $0.73^{+0.05}_{-0.10}$ & $0.66^{+0.03}_{-0.13}$ & $0.81^{+0.13}_{-0.11}$ & $0.78^{+0.01}_{-0.01}$ \\ 
$\sigma(J/\psi+\chi_{c2})$ & $0.88^{+0.33}_{-0.24}$ & $1.10^{+0.01}_{-0.12}$ & $0.76^{+0.16}_{-0.43}$ & $1.13^{+0.05}_{-0.10}$ & $0.80^{+0.13}_{-0.22}$ \\
$\sigma(\eta_{c}+\gamma)$ & $76.76$ & $64.19^{+1.82}_{-2.19}$ & $39.00^{+7.56}_{-11.00}$ & $65.31^{+1.52}_{-1.77}$ & $43.53^{+6.04}_{-8.15}$ \\ 
$\sigma(\chi_{c0}+\gamma)$ & $1.01$ & $1.15^{+0.02}_{-0.02}$ & $1.19^{+0.01}_{-0.01}$ & $1.16^{+0.03}_{-0.02}$ & $1.19^{+0.01}_{-0.01}$ \\ 
$\sigma(\chi_{c1}+\gamma)$ & $13.63$ & $10.73^{+0.42}_{-0.51}$ & $8.18^{+0.78}_{-1.04}$ & $10.97^{+0.36}_{-0.41}$ & $8.69^{+0.66}_{-0.86}$ \\ 
$\sigma(\chi_{c2}+\gamma)$ & $6.16$ & $2.63^{+0.51}_{-0.61}$ & $0.73^{+0.60}_{-0.72}$ & $3.20^{+0.36}_{-0.40}$ & $1.64^{+0.43}_{-0.47}$ \\ 
\hline
\end{tabular}
}
\end{center}
\end{table}

The predicted cross sections up to NNLO in the two prescriptions are collected in Table~\ref{tab:cross:section1}. The quoted uncertainties arise from varying the renormalization scale $\mu_R$ between $3$ GeV and $\sqrt{s}$, with the central value chosen as $\sqrt{s}/2$. As can be seen from Table~\ref{tab:cross:section1}, the predicted cross sections receive sizable higher-order corrections in both prescriptions. To assess the convergence of the perturbative series, we compare the NLO and NNLO predictions with the corresponding LO results for each channel. For the double-charmonium channels, the NLO corrections are positive. The most pronounced enhancement occurs in $e^{+}e^{-}\to J/\psi+\eta_{c}$, where the NLO prediction is about $2.3$ ($2.8$) times the LO result in the squared-amplitude-level (amplitude-level) prescription, increasing further to about $3.3$ ($3.7$) times the LO result at NNLO. For the $P$-wave channel $e^{+}e^{-}\to J/\psi+\chi_{c0}$, the NLO and NNLO predictions are about $1.9$ ($2.1$) and $2.3$ ($2.4$) times the LO result in the squared-amplitude-level (amplitude-level) prescription, respectively. For $e^{+}e^{-}\to J/\psi+\chi_{c1}$ and $e^{+}e^{-}\to J/\psi+\chi_{c2}$, the perturbative series exhibits an oscillatory behavior. The NLO predictions are about $1.4$ ($1.5$) and $1.3$ ($1.3$) times the LO results, respectively, while the NNLO predictions decrease to about $1.2$ ($1.4$) and $0.9$ ($0.9$) times the LO results.

For the photon-associated channels, the higher-order corrections are negative and systematically reduce the cross sections. For $e^{+}e^{-}\to\eta_{c}+\gamma$, $e^{+}e^{-}\to\chi_{c1}+\gamma$, and $e^{+}e^{-}\to\chi_{c2}+\gamma$, the NLO (NNLO) predictions are about $84\%$ ($51\%$), $79\%$ ($60\%$), and $43\%$ ($12\%$) of the LO results in the squared-amplitude-level prescription, respectively. In the amplitude-level prescription, the corresponding values are about $85\%$ ($57\%$), $80\%$ ($64\%$), and $52\%$ ($27\%$), respectively. In particular, for $e^{+}e^{-}\to\chi_{c2}+\gamma$, the NNLO$|_{\rm sq}$ cross section of $0.73$ fb is only about $12\%$ of the LO prediction, whereas the NNLO$|_{\rm amp}$ result of $1.64$ fb remains at about $27\%$ of the LO prediction. In contrast, the $e^{+}e^{-}\to\chi_{c0}+\gamma$ channel shows a much better-behaved perturbative expansion, with the NLO and NNLO predictions only about $14\%$--$15\%$ and $18\%$ above the LO result in both prescriptions.

It is noteworthy that, for every channel considered here, the NNLO/LO ratio is closer to the corresponding NLO/LO ratio in the amplitude-level prescription than in the squared-amplitude-level prescription. This indicates that the NNLO corrections are generally more moderate in the amplitude-level prescription, leading to a more stable and better-converged perturbative expansion.

In Table~\ref{tab:cross:section1}, we also compare the renormalization scale uncertainties of the NNLO predictions obtained in the two prescriptions. Here, renormalization scale $\mu_R$ is varied between $3$ GeV and $\sqrt{s}$, with the central value set to $\sqrt{s}/2$. It can be seen that the amplitude-level prescription considerably reduces the residual scale dependence in every channel. For $e^{+}e^{-}\to J/\psi+\eta_{c}$, the scale uncertainty of the NNLO cross section decreases from $^{+3.82}_{-2.96}$ fb in the squared-amplitude-level prescription to $^{+2.33}_{-2.40}$ fb in the amplitude-level prescription, corresponding to a reduction in the relative scale uncertainty from about $^{+25\%}_{-19\%}$ to $^{+14\%}_{-14\%}$. For $e^{+}e^{-}\to J/\psi+\chi_{c0}$, the scale uncertainty decreases from $^{+0.45}_{-0.71}$ fb to $^{+0.15}_{-0.42}$ fb, corresponding to a reduction in the relative uncertainty from about $^{+7\%}_{-12\%}$ to $^{+2\%}_{-7\%}$. For $e^{+}e^{-}\to J/\psi+\chi_{c1}$ and $e^{+}e^{-}\to J/\psi+\chi_{c2}$, the relative scale uncertainties decrease from about $^{+5\%}_{-20\%}$ and $^{+21\%}_{-57\%}$ to about $^{+1\%}_{-1\%}$ and $^{+16\%}_{-28\%}$, respectively.

For the photon-associated channels, a reduction in the scale dependence is also observed. For $e^{+}e^{-}\to\eta_{c}+\gamma$, the relative scale uncertainty decreases from about $^{+19\%}_{-28\%}$ to $^{+14\%}_{-19\%}$, while for $e^{+}e^{-}\to\chi_{c0}+\gamma$, it remains at the level of about $1\%$ in both prescriptions. For $e^{+}e^{-}\to\chi_{c1}+\gamma$, the relative scale uncertainty decreases from about $^{+10\%}_{-13\%}$ to $^{+8\%}_{-10\%}$. The improvement is most pronounced for $e^{+}e^{-}\to\chi_{c2}+\gamma$, where the relative scale uncertainty of the NNLO$|_{\rm sq}$ prediction is as large as $^{+82\%}_{-99\%}$ but is reduced to $^{+26\%}_{-29\%}$ in the amplitude-level prescription. These findings are consistent with the $\mu_R$ dependence shown in Fig.~\ref{urdependence}, indicating that the amplitude-level prescription reduces the residual renormalization-scale dependence of the NNLO predictions.

\begin{table*}[t]
\small
\begin{center}
\caption{NNLO cross sections (in units of fb) of $e^{+}e^{-}\to J/\psi+\eta_{c}$, $e^{+}e^{-}\to J/\psi+\chi_{cJ}$, $e^{+}e^{-}\to \eta_{c}+ \gamma$, and $e^{+}e^{-}\to \chi_{cJ}+ \gamma$ $(J=0,1,2)$ in the squared-amplitude-level prescription and the amplitude-level prescription. The corresponding experimental measurements from the Belle and BaBar collaborations are also included for comparison. The first uncertainties originate from the charm quark mass $m_c = 1.5 \pm 0.2$ GeV, and the second from varying the renormalization scale $\mu_R$ between $3$ GeV and $\sqrt{s}$, with the central value set to $\sqrt{s}/2$.}
\label{tab:cross:section2}
\resizebox{0.85\textwidth}{!}{
\begin{tabular}{|c|c|c|c|c|}
\hline
 & NNLO$|_{\rm sq}$ & NNLO$|_{\rm amp}$ & {\rm Belle}~\cite{Belle:2004abn, Belle:2018jqa}& {\rm BaBar}~\cite{BaBar:2005nic}\\ \hline
$\sigma(J/\psi+\eta_{c})$ & $15.31^{+7.69+3.82}_{-4.58-2.96}$ & $17.22^{+9.51+2.33}_{-5.45-2.40}$ & $25.6\pm2.8\pm3.4$ & $17.6\pm2.8^{+1.5}_{-2.1}$\\  
$\sigma(J/\psi+\chi_{c0})$ & $6.09^{+6.66+0.45}_{-2.95-0.71}$ & $6.30^{+7.09+0.15}_{-3.09-0.42}$ & $6.4\pm1.7\pm1.0$ & $10.3\pm2.5^{+1.4}_{-1.8}$\\  
\parbox{3cm}{\centering $\sigma(J/\psi+\chi_{c1})$\\+$\sigma(J/\psi+\chi_{c2})$} & $1.42^{+2.12+0.19}_{-0.84-0.56}$ & $1.58^{+2.35+0.14}_{-0.92-0.23}$ & $<5.3$ at $90\%$ C.L.  & $-$ \\
$\sigma(\eta_{c}+\gamma)$ & $39.00^{+9.82+7.56}_{-7.08-11.00}$ & $43.53^{+12.20+6.04}_{-8.51-8.15}$ & $<21.1$ at $90\%$ C.L. & $-$\\ 
$\sigma(\chi_{c0}+\gamma)$ & $1.19^{+1.07+0.01}_{-0.53-0.01}$ & $1.19^{+1.06+0.01}_{-0.53-0.01}$ & $<205.9$ at $90\%$ C.L. & $-$ \\ 
$\sigma(\chi_{c1}+\gamma)$ & $8.18^{+5.58+0.78}_{-2.91-1.04}$ & $8.69^{+5.93+0.66}_{-3.09-0.86}$ & $17.3^{+4.2}_{-3.9}\pm1.7$ & $-$\\ 
$\sigma(\chi_{c2}+\gamma)$ & $0.73^{+0.26+0.60}_{-0.16-0.72}$ & $1.64^{+0.88+0.43}_{-0.48-0.47}$ & $<5.7$ at $90\%$ C.L. & $-$\\ \hline
\end{tabular}
}
\end{center}
\end{table*}

In Table~\ref{tab:cross:section2}, the NNLO predictions obtained with the two prescriptions are compared with the corresponding experimental measurements from the Belle~\cite{Belle:2004abn, Belle:2018jqa} and BaBar~\cite{BaBar:2005nic} collaborations. The first uncertainties originate from varying the charm quark mass, $m_c=1.5\pm0.2$ GeV, while the second arise from the renormalization scale variation. In the channels where the cross sections have been measured, the amplitude-level prescription gives central values that are closer to the data. For $e^{+}e^{-}\to J/\psi+\eta_{c}$, the amplitude-level prediction of $17.22$ fb almost coincides with the BaBar measurement of $17.6$ fb and is consistent with the Belle measurement of $25.6$ fb within the theoretical and experimental uncertainties, whereas the squared-amplitude-level prediction of $15.31$ fb is about $13\%$ lower than the BaBar central value. For $e^{+}e^{-}\to J/\psi+\chi_{c0}$, the amplitude-level prediction of $6.30$ fb agrees well with the Belle measurement of $6.4$ fb, and the predictions of both prescriptions are also compatible with the BaBar measurement of $10.3$ fb within the uncertainties. For $e^{+}e^{-}\to\chi_{c1}+\gamma$, the Belle collaboration has measured a cross section of $17.3^{+4.2}_{-3.9}\pm1.7$ fb; the NNLO predictions of $8.18$ fb and $8.69$ fb in the two prescriptions lie below the central value but remain consistent with the data once the sizable theoretical and experimental uncertainties are taken into account. In the channels where only upper limits are currently available, the predictions of the two prescriptions are consistent with the limits. For $J/\psi+\chi_{c1}$ and $J/\psi+\chi_{c2}$, the combined NNLO cross sections of $1.42$ fb and $1.58$ fb are well below the Belle upper limit of $5.3$ fb. For $e^{+}e^{-}\to\chi_{c0}+\gamma$, the NNLO predictions of $1.19$ fb in both prescriptions are more than two orders of magnitude below the Belle upper limit of $205.9$ fb. For $e^{+}e^{-}\to\chi_{c2}+\gamma$, the NNLO predictions of $0.73$ fb and $1.64$ fb are also well below the Belle upper limit of $5.7$ fb. However, the NNLO predictions for $e^{+}e^{-}\to\eta_{c}+\gamma$ of both prescriptions, $39.00$ fb in the squared-amplitude-level prescription and $43.53$ fb in the amplitude-level prescription, still exceed the Belle upper limit of $21.1$ fb, although the NNLO corrections substantially reduce the cross section compared with the lower orders.

\section{Summary} \label{III}

In summary, we have analyzed the NNLO QCD corrections to the exclusive processes $e^{+}e^{-}\to J/\psi+\eta_{c}$, $e^{+}e^{-}\to J/\psi+\chi_{cJ}$, $e^{+}e^{-}\to \eta_{c}+\gamma$, and $e^{+}e^{-}\to \chi_{cJ}+\gamma$ $(J=0,1,2)$ at the $B$ factories within the squared-amplitude-level and amplitude-level prescriptions. 

Our numerical results show that the amplitude-level prescription leads to better perturbative convergence than the conventional squared-amplitude-level prescription. In all channels studied, the NNLO/LO ratios in the amplitude-level prescription lie closer to the corresponding NLO/LO ratios, indicating more moderate NNLO corrections. This improvement is particularly pronounced for channels with poor perturbative convergence, such as $e^{+}e^{-}\to\chi_{c2}+\gamma$. Moreover, the amplitude-level prescription reduces the residual renormalization-scale dependence of the NNLO predictions, with the most pronounced improvement occurring in the channel that exhibit the strongest scale dependence in the squared-amplitude-level prescription. This results in more stable NNLO predictions under variations of the renormalization scale. The NNLO predictions in the amplitude-level prescription are in good agreement with the available data and have central values closer to the measurements than those in the squared-amplitude-level prescription. For channels with only upper limits, both prescriptions are consistent with the experimental constraints, except for $e^{+}e^{-}\to\eta_c+\gamma$, where both NNLO predictions remain above the current upper limit. These features make the amplitude-level prescription a more robust framework for NNLO predictions of exclusive heavy-quarkonium production at the $B$ factories. Future measurements of additional processes will provide further tests of this prescription.

\hspace{2cm}

\noindent {\bf Acknowledgments:} We would like to thank Cong Li for useful discussions. The work is supported by the National Natural Science Foundation of China under Grant Nos. 12505097, 12135013 and the Chongqing Natural Science Foundation under Grant No. CSTB2025NSCQ-GPX1018.

\hspace{2cm}

\end{document}